**Polarization gratings spontaneously formed by the helical Twist-Bend Nematic phase**

*N. Vaupotič, M. Ali, P. W. Majewski, E. Gorecka, D. Pociecha\**

Prof. N. Vaupotič
Department of Physics, Faculty of Natural Sciences and Mathematics, University of Maribor, Koroška 160, 2000 Maribor, Slovenia
Jožef Stefan Institute, Jamova 39, 1000 Ljubljana, Slovenia

M. Ali, Dr. P. Majewski, Prof. E. Gorecka, Dr. D. Pociecha
Faculty of Chemistry, University of Warsaw
ul. Zwirki i Wigury 101, 02-089 Warsaw, Poland
E-mail: pociu@chem.uw.edu.pl

Spontaneous formation of polarization gratings by liquid crystals made of bent dimeric molecules is reported. The grating is formed within the temperature range of the twist bend modulated nematic phase, $N_{TB}$, without the necessity to pattern the cell surfaces, therefore the modulated nematic phase is a promising candidate for low-cost modulators and beam steering devices, the polarization properties of which can be tuned by temperature. In addition, the study of the diffracted light properties turns out to be a sensitive measuring technique for determination of the 3D spatial variation of the optic axis in the cell.

**1. Introduction**
Diffraction gratings formed by anisotropic media with a spatial variation of the optic axis direction enable amplitude, phase and polarization modulation of the diffracted light. Liquid crystals have been extensively studied in this respect, due to their high birefringence, which is easily tunable by external fields. The most studied are polarization gratings formed by the nematic liquid crystal phase, in which spatial variation of the optic axis is achieved through patterning the surfaces by photolithography,[1] laser scanning,[2] mask photopolymerization,[3] microrubbing,[4] nanorubbing,[5] azo-dye command layer,[6] chemisorption,[7] photoalignment and holographic recording,[8] and periodic electrode structures[9]. A thorough overview of the preparation methods for optic axis gratings can be found in Ref. [10]. Diffraction gratings have been prepared also from polymer and polymer dispersed liquid crystals.[11,12] Photopolymerization of liquid crystalline phases has been proposed as an easy tool to fabricate gratings with electrically controllable diffraction efficiency.[13] Of special importance are liquid crystalline phases that spontaneously form space-modulated structures, because no surface patterning is required to produce the grating. Cholesteric and chiral smectic phases have been considered in this respect.[14,15]

Recently, a tunable optical grating made of flexoelectric domains in a bent-core nematic liquid crystal has been reported.[16] Bent-core liquid crystals have very unique physical properties, which arise from the specific shape of the constituent molecules.[17] They form several intriguing phases, one of them, the recently discovered twist-bend nematic phase, $N_{TB}$, is a unique example of a system with spontaneous chiral symmetry breaking.[18] Although it is formed by achiral molecules, bent-core or odd-linkage dimers, the phase has a heliconical molecular arrangement with an extremely short pitch of 8-15 nm, corresponding to just a few molecular distances. The commonly accepted model of the $N_{TB}$ phase assumes that molecules are inclined from the heliconical axis at an arbitrary conical angle and process on the cone. Because the pitch in the $N_{TB}$ phase is much shorter than the wavelength of visible light, the phase is optically uniaxial, with the optic axis along the helix, and the ordinary and extraordinary refractive indices dependent on the conical angle and therefore on temperature.[19] In thin cells with planar anchoring, a structure with periodic spatial variation of the local heliconical axis direction, therefore optical indicatrix, is formed spontaneously at temperatures below the N-$N_{TB}$ phase transition; the period of the modulation is always twice the cell thickness. Such a pattern effectively diffracts light and changes the polarization of the diffracted beams. In this report, we show that the analysis of the polarization state of the diffracted light reveals details of the spatial variation of the heliconical axis. Thus, the study of the diffraction pattern is a sensitive experimental technique to determine the internal structure of the cell. The second order diffraction peaks of the incident circularly polarised light turned out to be the most sensitive to the temperature induced structure changes.

## 2. Results and Discussion

The studied material is a symmetric dimer CB7CB, the model $N_{TB}$ compound. Upon cooling, CB7CB exhibits transition from the isotropic liquid to the nematic phase at 114 °C and from the nematic to the twist-bent nematic phase at 102 °C. The material in the isotropic liquid phase was filled by the capillary action into a glass cell (1.6 μm thick) with polymer aligning layers and unidirectional rubbing, ensuring a uniform planar alignment in the nematic phase. The experimental geometry is given in **Figure 1**. The cell surfaces are in the $xy$-plane and light propagates along the $z$-direction through the cell of thickness $L$. The grating modulation is along the $x$-direction, its periodicity is $p$. The diffraction of a linearly and circularly polarised light was studied by monitoring the azimuthal angle $\psi$ (the inclination of the long ellipse axis) and ellipticity $e$ (the arctan of the ratio between the short and long ellipse axis) of the diffracted light. The incident linear polarization (LP) is called horizontal (HLP, azimuthal angle $\psi = 0$) for the light polarized along the $x$-axis and vertical (VLP, azimuthal angle $\psi = 90°$) when polarized along the $y$-axis. The diffracted light is, in general, elliptically polarized. Far below the N-$N_{TB}$ phase transition the average phase retardation for the 1.6-μm thick sample and 633 nm wavelength is approximately $\pi/2$ and thus the cell behaves as a quarter-wave plate for the direct beam: the outgoing direct beam is almost linearly polarized for the circularly polarized (CP) incident light and vice versa (**Table 1**). For the circularly polarized incident light the first order diffraction peaks, observed at $q = \pm q_0$ with $q_0 = 2\pi/p$, are circularly polarized but with a reversed circularity with respect to the incident light, i.e. for the left-handed CP incident light the first order diffraction peaks are right-handed CP. For the linearly polarized incident light, the $\pm q_0$ beams are nearly linearly polarized (ellipticity is below 2°), with the $x$ and $y$ components of the electric field exchanged with respect to the incident light, i.e. the horizontally/vertically polarized light is transformed into vertically/horizontally polarized, while the polarization plane of the diagonally polarized incident light is not affected (Table 1). The second order diffraction peaks ($q = \pm 2q_0$) are elliptically polarized for LP and CP incident light, with a small asymmetry between the $+2q_0$ and $-2q_0$ peaks. For the circularly polarized incident light the second order diffraction peaks have a small ellipticity, below 10°, while in case of the linearly polarized incident light the ellipticity of the diffracted beams strongly depends on the polarization plane of the incident light, being maximal (~ 30°) for the diagonal polarization direction.

To account for the observed properties of the diffracted light we consider different one-dimensional spatial variations of the optic axis. Let the direction ($\vec{n}$) of the optic axis be given as: $\vec{n} = (\sin\alpha \cos\beta, \cos\alpha \cos\beta, \sin\beta)$ where $\alpha$ and $\beta$ vary periodically along $x$ with periodicity $p$. We have tested and compared the results of the following modulations:

- Model 1: sinusoidal in-plane modulation: $\alpha = \alpha_0 \sin\left(\frac{2\pi}{p}x\right)$ and $\beta = 0$.

- Model 2: sinusoidal out-of-plane modulations, with a relative shift by $\frac{\pi}{2}$ between $\alpha$ and $\beta$: $\alpha = \alpha_0 \sin\left(\frac{2\pi}{p}x\right)$ and $\beta = \beta_0 \cos\left(\frac{2\pi}{p}x\right)$.

- Model 3: sinusoidal out-of-plane modulations, with an arbitrary relative shift ($s$) of $\beta$ with respect to $\alpha$: $\alpha = \alpha_0 \sin\left(\frac{2\pi}{p}x\right)$ and $\beta = \beta_0 \cos\left(\frac{2\pi}{p}(x+s)\right)$.

The $x$ and $y$ components of the electric field ($A_x^{(out)}$ and $A_y^{(out)}$) in the transmitted light are calculated as:

$$\begin{pmatrix} A_x^{(out)} \\ A_y^{(out)} \end{pmatrix} = T \begin{pmatrix} A_x^{(in)} \\ A_y^{(in)} \end{pmatrix},$$

where $A_x^{(in)}$ and $A_y^{(in)}$ are the electric field components of the incident light,

$$T = \begin{pmatrix} \cos\alpha & \sin\alpha \\ -\sin\alpha & \cos\alpha \end{pmatrix} \cdot \begin{pmatrix} 1 & 0 \\ 0 & e^{i\phi} \end{pmatrix} \cdot \begin{pmatrix} \cos\alpha & -\sin\alpha \\ \sin\alpha & \cos\alpha \end{pmatrix}$$

$$= \begin{pmatrix} 1+\sin^2\alpha & -\sin\alpha\cos\alpha \\ -\sin\alpha\cos\alpha & 1+\cos^2\alpha \end{pmatrix}(e^{i\phi}-1)$$

is the transfer matrix and $\phi = 2k_0\Delta nL$ is the phase difference between the ordinary and extraordinary components of the electric field. The birefringence $\Delta n$ depends on the angle $\beta$:

$$\Delta n = \sqrt{\left(\frac{\sin^2\beta}{n_o^2} + \frac{\cos^2\beta}{n_e^2}\right)^{-1}} - n_o \qquad (1)$$

where $n_o$ and $n_e$ are the ordinary and extraordinary indices of refraction, respectively. The electric field components ($A_x^{(q)}$ and $A_y^{(q)}$) of the light diffracted at a given diffraction wave vector $q$ are:

$$\begin{pmatrix} A_x^{(q)} \\ A_y^{(q)} \end{pmatrix} = \left(\int_0^p \begin{pmatrix} A_x^{(out)} \\ A_y^{(out)} \end{pmatrix} e^{iqx} dx\right)\left(\sum_{j=0}^{N-1} e^{iqpj}\right) \qquad (2)$$

where $N$ is the number of illuminated "units" (by a "unit" we mean the length of one full modulation of the optic axis). If several "units" are illuminated, the sum in Equation (2) becomes a delta function, which differs from zero only for multiples of $q_0$. To obtain the electric field in the diffracted light one thus has to calculate the Fourier transform of the transfer matrix. In general, the $x$ and $y$ components of the electric field at a given $nq_0$ are complex numbers, their magnitudes and the difference between their phases define the polarization state of the diffracted light (for details of the calculation see the Supporting Information). In general, the outgoing light is elliptically polarized. The electric field components ($E_x^{(q)}$ and $E_y^{(q)}$) in the outgoing light can be expressed as:

$$\begin{pmatrix} E_x^{(q)} \\ E_y^{(q)} \end{pmatrix} = \begin{pmatrix} \left|A_x^{(q)}\right|\cos(\varphi_x - \omega t) \\ \left|A_y^{(q)}\right|\cos(\varphi_y - \omega t) \end{pmatrix}$$

where the dependence on time ($t$) was added and $\omega = 2\pi\nu$, where $\nu$ is the frequency of light.

The orientation of the long ellipse axis ($\psi$) is obtained as $\psi = \tan^{-1}\left(E_y^{(q)}/E_x^{(q)}\right)$ at $t$, at which the magnitude of the electric field in the diffracted light is maximum.

In **Table 2** we give the orientation of the long ellipse axis and ellipticity of the diffracted light obtained by each model for different polarization states of the incident light. The ellipticity for the left rotation is denoted by negative values and for the right rotation by positive values. The values of parameters chosen for the calculations are: $\alpha_0 = 30°$, $\beta_0 = 30°$ (when applicable), $p = 3.2\ \mu m$, $L = 1.6\ \mu m$ and $s = 0.02p$ (when applicable). For the value of the ordinary and extraordinary indices of refraction we choose $n_e = 1.6$ and $n_o = 1.5$, respectively. With these parameters, the cell with no modulation of optic axis behaves as a quarter-wave plate for the direct beam, as observed experimentally.

The model 1 with only the in-plane variation of the optic axis is clearly inconsistent with the experimental results because it predicts that the ellipticity of light is not changed neither for $q_0$ nor $2q_0$ diffracted beams, i.e. the linearly/circularly polarized incoming light is transformed to linearly/circularly polarized outgoing beams. The predictions of model 2, in which the in-plane ($\alpha$) and out-of-plane ($\beta$) modulations of optic axis are of equal amplitude but shifted by a quarter of the modulation wavelength, are closer to the experimental results. This model shows that for the $\pm 2q_0$ beams the diagonally polarized incoming light is transformed into the elliptically polarized ($e \sim 33°$) outgoing light and circularly polarized into linearly polarized. A good agreement between the experiment and model 2 is obtained if deep in the $N_{TB}$ phase the amplitudes of $\alpha$ and $\beta$ are close to 30°, in agreement with experimental value of conical angle deduced from birefringence measurements (see SI). However, model 2 does not explain the asymmetry of the $\pm 2q_0$ diffracted beams. Model 3, which includes some shift between the regions where $\alpha$ and $\beta$ are maximum, explains also the asymmetry. It should be pointed out, that the required asymmetry is very low (only $0.02p$).

What drives the observed modulations of the optic axis? First, it should be noticed that due to the heliconical structure, the $N_{TB}$ phase can be considered as a pseudo-layered medium. In the temperature range just below the N-$N_{TB}$ phase transition, in which the heliconical angle changes rapidly, the pseudo-layers shrink. Under the constant density condition, this causes instability and the layers start to undulate, both in the horizontal ($xy$) and vertical ($xz$) planes (**Figure 2**). The horizontal undulations cause the rotation of the optic axis by the angle $\alpha$, while the vertical undulations rotate the optic axis from the surface by the angle $\beta$ and are responsible for the decrease of birefringence in some areas. In thin cells such undulations are smooth, leading to a stripe texture that is characteristic for the $N_{TB}$ phase, with the stripe periodicity equal to the cell thickness. In thick cells the strongly curved layers become localized in some areas, while in other areas the layers are almost straight, leading to a formation of an array of focal conics at the cell surface, instead of a stripe texture (**Figure S3**).

In order to determine the deformation of the pseudo-layered structure with temperature we also checked the temperature dependence of the polarization state of the diffracted light. For circularly polarized incident light, it was observed that deep in the $N_{TB}$ phase the $2q_0$ beams are nearly linearly polarized but on approaching the transition temperature to the nematic phase the polarization of the $2q_0$ beams changes dramatically, becoming nearly circular (**Figure 3**). Heating the sample reduces the heliconical angle in the $N_{TB}$ phase and therefore increases the birefringence. However, such changes should affect mainly the azimuthal angle $\psi$ and should have only a minor effect on the ellipticity of the $2q_0$ diffracted beams (Figure 3). Therefore, to account for the observed changes of the $2q_0$ diffracted beams polarization one has to assume a change in the type of modulation as well. Deep in the $N_{TB}$ phase, a good agreement with the experimental results is obtained by model 3, if the amplitudes of angles $\alpha$ and $\beta$ are taken to be equal. To account for the observed changes of ellipticity of the diffracted light we have to release this constraint. Interestingly, even if the asymmetry is very small, e.g. $\beta$ is only 10% smaller than α, the ellipticity of the $2q_0$ diffracted beam increases significantly (see Figure 3(b),(c) and **Table S1**). By reducing $\beta$ to zero the results converge to those predicted by model 1. This clearly shows that the pseudo-layers start to stretch vertically as temperature increases while the amplitude of the horizontal modulation is less affected.

## 3. Conclusions

We showed that the modulated structure that is spontaneously produced in thin cells of the $N_{TB}$ phase can serve as a polarization grating, transforming a linearly polarized incident light into circularly polarized diffracted light and vice versa. The importance of this result is in the fact that no expensive procedures for the surface patterning are required to produce the grating. We focused primarily on the possibility for the diffraction on such a grating to serve as a simple and inexpensive experimental tool to study the structure of the material in a confined geometry, namely the spatial variation of the optic axis. This information can be obtained by a detailed analysis of the polarization state of the diffracted beams. We showed that for the CB7CB material far from the N-$N_{TB}$ phase transition temperature the pseudo-layer structure of the $N_{TB}$ phase has similar amplitudes of the horizontal and vertical undulations. The increase in temperature stretches the vertical undulation. The asymmetry in the undulation amplitude strongly influences the grating properties and as a result, the polarization of the diffracted beams changes significantly.

**Acknowledgements**

This work was supported by the National Science Centre (Poland) under Grant No. 2015/19/P/ST5/03813. NV acknowledges the support of the Slovenian Research Agency (ARRS), through the research project P1-0055.

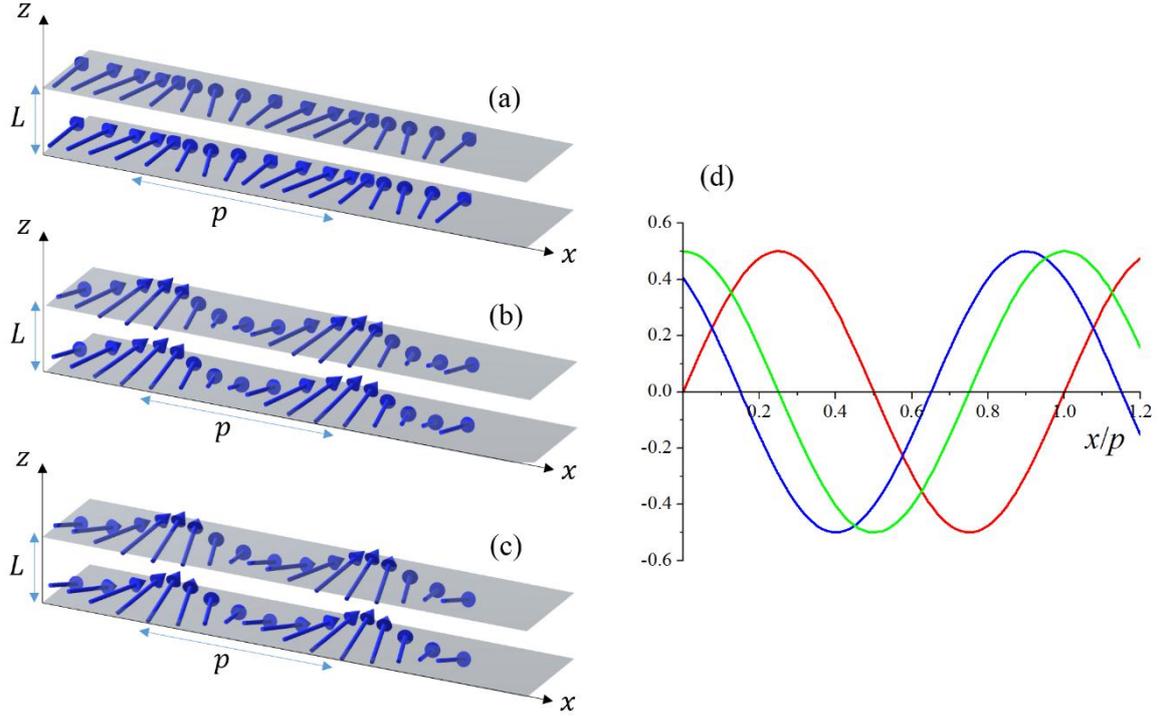

**Figure 1.** Spatial variation of the optic axis in a) model 1, b) model 2 and c) model 3. d) Spatial dependence of the in-plane orientation of the direction of optic axis, α, for model 1 (red) and the out-of-plane tilt by $\beta$: model 2 (green,) and model 3 with shift $s = 0.1p$ (blue).

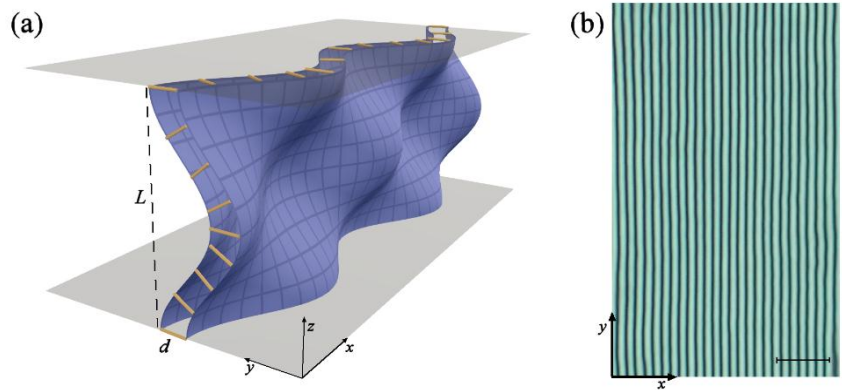

**Figure 2.** a) 3D undulations of the pseudo-layers of thickness $d$ (helical pitch) in the thin cell of thickness $L$ result in the b) stripe texture that acts as a diffraction grating. Yellow bars in (a) represent the direction of the optic axis that is along the heliconical axis in the $N_{TB}$ phase; the scale bar in (b) corresponds to 10 μm.

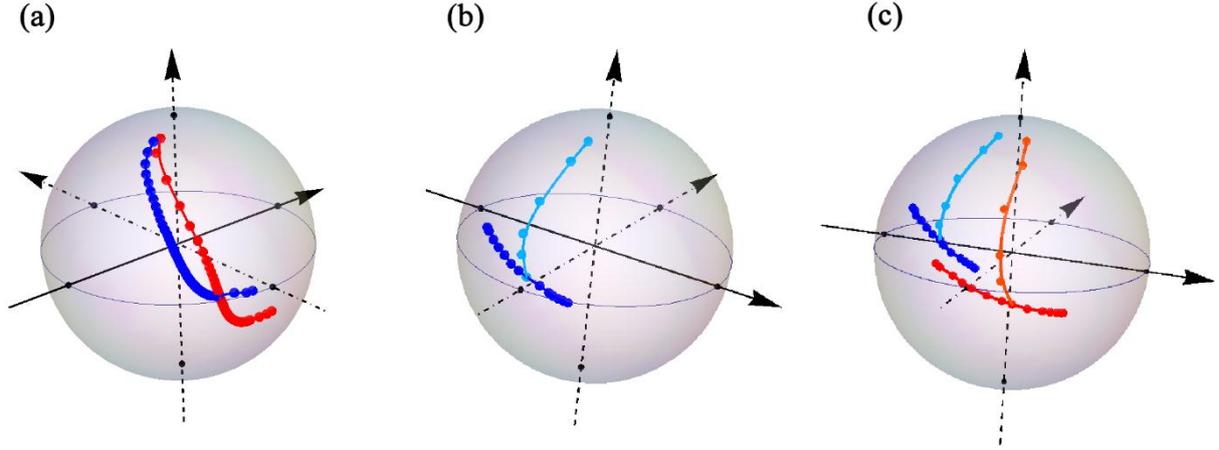

**Figure 3.** Measured and theoretical normalized values of the Stokes parameters ($s_1, s_2, s_3$), presented on the Poincaré sphere; solid arrow: $s_1 = \cos(2\psi)\cos(2e)$; dash-dotted arrow: $s_2 = \sin(2\psi)\cos(2e)$; dashed arrow: $s_3 = \sin(2e)$. a) Experimental temperature dependence of the polarization state of the $+2q_0$ (red) and $-2q_0$ (blue) diffracted beams for the left circularly polarized incident light. b) Temperature development of the polarization state as obtained from model 2 for $\alpha_0 = \beta_0$ (blue) and for $\alpha_0 > \beta_0$ (cyan). c) Temperature development of the polarization state as obtained from model 3 for $\alpha_0 = \beta_0$ (red: $+2q_0$ peak, blue: $-2q_0$ peak) and for $\alpha_0 > \beta_0$ (orange: $+2q_0$ peak, cyan: $-2q_0$ peak). Values of the parameters ($\alpha_0, \beta_0$ and birefringence) as well as the calculated values of the azimuthal angle $\psi$ and ellipticity $e$ are given in the Supporting Information.

**Table 1**. Experimental data. Azimuthal angle of the long ellipse axis ($\psi$) and ellipticity ($e$), both given in degrees, of the diffracted light for the horizontal linear (HLP), vertical linear (VLP), diagonal linear (DLP) and left circular (LCP) polarized incident light with wavelength 633 nm. The values for the direct beam and the first and second order diffraction peaks are given; $q_0 = 2\pi/p$, where $p$ is the wavelength of modulation. The cell thickness at 60 °C is 1.65 μm.

|  | HLP | | VLP | | DLP | | LCP | |
|---|---|---|---|---|---|---|---|---|
| $q$ | $\psi$ | $e$ | $\psi$ | $e$ | $\psi$ | $e$ | $\psi$ | $e$ |
| $2q_0$ | −5.5 | 1 | −88 | −4.5 | 22 | −39 | −45 | −5 |
| $q_0$ | 87 | 2 | −2.5 | 0 | 43 | −1.5 | / | 42 |
| 0 | −1.5 | −1.5 | 88 | −2.8 | / | −43 | −44 | −3 |
| $-q_0$ | −87 | −1 | −3 | 0 | 41 | −2 | / | 43 |
| $-2q_0$ | 2 | 5 | −88 | 13 | 52 | −35 | −39 | 3 |

**Table 2**. Theoretically calculated azimuthal angle of the long ellipse axis ($\psi$) and ellipticity ($e$) for all four models. Symbols and abbreviations are defined in the text and in Table 1.

| model | | HLP | | VLP | | DLP | | LCP | |
|---|---|---|---|---|---|---|---|---|---|
| | $q$ | $\psi$ | $e$ | $\psi$ | $e$ | $\psi$ | $e$ | $\psi$ | $e$ |
| 1 | $2q_0$ | 0 | 0 | 90 | 0 | −45 | 0 | / | 45 |
| | $q_0$ | 90 | 0 | 0 | 0 | 45 | 0 | / | 45 |
| | 0 | 0 | 0 | 90 | 0 | 45 | −37 | −45 | 8 |
| | $-q_0$ | 90 | 0 | 0 | 0 | 45 | 0 | / | 45 |
| | $-2q_0$ | 0 | 0 | 90 | 0 | −45 | 0 | / | 45 |
| 2 | $2q_0$ | 0 | 0 | 90 | 0 | −1 | −33 | −33 | 0 |
| | $q_0$ | 90 | 0 | 0 | 0 | 45 | 0 | / | 45 |
| | 0 | 0 | 0 | 90 | 0 | 45 | −31 | −45 | −14 |
| | $-q_0$ | 90 | 0 | 0 | 0 | 45 | 0 | / | 45 |
| | $-2q_0$ | 0 | 0 | 90 | 0 | −1 | −33 | −33 | 0 |
| 3 | $2q_0$ | 0 | 0 | 90 | 0 | 6 | −22 | −23 | −4 |
| | $q_0$ | 90 | 0 | 0 | 0 | 45 | 0 | / | 45 |
| | 0 | 0 | 0 | 90 | 0 | 45 | −31 | −45 | −14 |
| | $-q_0$ | 90 | 0 | 0 | 0 | 45 | 0 | / | 45 |
| | $-2q_0$ | 0 | 0 | 90 | 0 | −28 | −39 | −42 | 5 |

# Supporting Information

**Polarization gratings spontaneously formed by the helical Twist-Bend Nematic phase**

*Nataša Vaupotič, Muhammad Ali, Pawel W. Majewski, Ewa Gorecka, Damian Pociecha\**

**Experimental Section**

Optical birefringence was measured by the photoelastic modulator (Hinds, PEM-90) setup. As a light source, a halogen lamp equipped with a narrow band pass filter (532 nm and 633 nm) was used. The conical tilt angle ($\theta$) in the twist-bend nematic phase ($N_{TB}$) was deduced from the decrease of the birefringence ($\Delta n$) with respect to the values measured in the nematic (N) phase, $\Delta n_{N_{TB}} = \Delta n_N (3\cos^2\theta - 1)/2$. The birefringence of the nematic phase was extrapolated to the lower temperature range by assuming a power law temperature dependence: $\Delta n = \Delta n_0 (T_c - T)^\gamma$, where $\Delta n_0$, $T_c$, and $\gamma$ are the fitting parameters. The diffraction experiments were performed with the HeNe laser (633 nm) and the polarization state of the incident and diffracted beams was analyzed by the PAX-1000 polarimeter (ThorLabs). The optical textures were observed by the polarizing microscope Zeiss AxioImager A2m. In all optical experiments, the temperature of the sample was stabilized with a precision of 0.1 K using the Mettler Toledo FP82HT heating stage. The AFM images were taken by the Bruker Dimension Icon microscope, working in the tapping mode at the liquid crystalline-air surface. Cantilevers with a low spring constant, $k = 0.4$ Nm$^{-1}$, were used. The resonant frequency was in a range between $70 - 80$ kHz and the typical scan frequency was 1 Hz.

**Additional results**

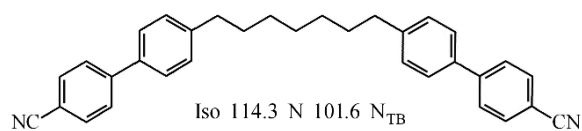

**Figure S1**. Molecular structure of the studied compound CB7CB and the transition temperatures detected on cooling.

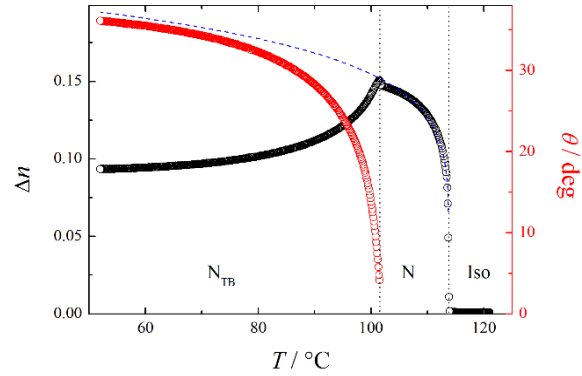

**Figure S2**. Temperature dependence of the optical birefringence (black circles) of the CB7CB material measured with red light, $\lambda = 633$ nm, and the conical tilt angle ($\theta$) in the twist-bend nematic phase (red circles). The dashed blue line shows the extrapolation of $\Delta n$ measured in the nematic phase to the temperature range of the $N_{TB}$ phase.

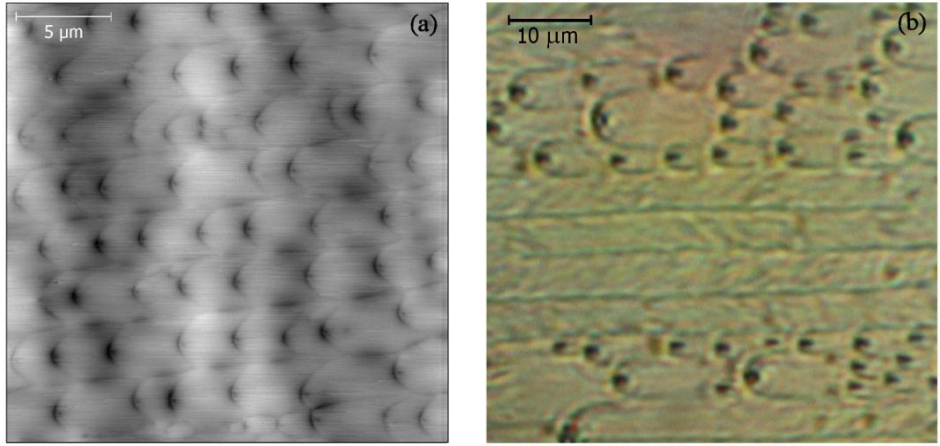

**Figure S3**. a) The AFM and b) optical image of the focal conics formed in the $N_{TB}$ phase of the CB7CB compound in the 3- and 5-μm thick cells.

**Details of theoretical calculations**

In calculating the phase difference between the $x$ and $y$ components of the diffracted light (equation (2) in the main text) there should be some caution in determining the phase difference between the two components. The phase of one component ( $\varphi_x = \tan^{-1}\left(\frac{\mathrm{Im}\left(A_x^{(q)}\right)}{\mathrm{Re}\left(A_x^{(q)}\right)}\right)$ or $\varphi_y = \tan^{-1}\left(\frac{\mathrm{Im}\left(A_y^{(q)}\right)}{\mathrm{Re}\left(A_y^{(q)}\right)}\right)$) should be defined by an angle between 0 and $2\pi$ measured from the real axis in the positive (anti-clock-wise) direction while the computer programs will give the values between $-\pi/2$ and $\pi/2$. Thus, if the phase is between 0 and $\pi/2$ and both the real and imaginary components are negative, one has to add $\pi$ to the calculated value of the phase. If the phase is between 0 and $-\pi/2$, then $\pi$ has to be added if the imaginary component is positive and the real component is negative. Failure to do so will give the wrong direction of the circularly polarized light and the wrong direction and the long axis orientation for the elliptically polarized light.

The parameters used in modelling the temperature dependence are shown in **Figure S4**. The amplitude of the modulation angle $\alpha$ ($\alpha_0$) is assumed to be equal to the heliconical angle. The amplitude of the modulation angle $\beta$ ($\beta_0$) is assumed to be equal to the $\alpha_0$, except close to the phase transition temperature (red points in Figure S4).

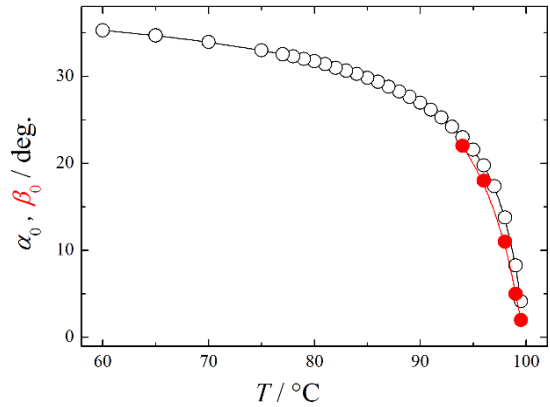

**Figure S4**. Temperature ($T$) dependence of the modulation angles amplitudes ($\alpha_0$, $\beta_0$) used in modelling.

Because an average birefringence ($\Delta n_{exp}$) is measured, we calculated the extraordinary ($n_e$) and ordinary ($n_o$) refractive indices entering equation (1) in the main text in such a way that the average birefringence in the model equals the measured birefringence. This was done by solving the set of equations:

$$n_{av}^2 = \frac{2n_o^2 + n_e^2}{3}$$

and

$$n_e = n_o + \Delta n_{exp} f \quad,$$

where the parameters $n_e, n_o$ and $f$ are found such as to match

$$\Delta n_{exp} = \frac{1}{p} \int_0^p \Delta n(\beta)\, dx \quad ,$$

where $\Delta n(\beta)$ is given by equation (1) in the main text.

The normalised Stokes parameters presenting the properties of the polarization of the diffracted light for the $\pm 2q_0$ peaks are given in the main text. Here we show the temperature dependence of the azimuthal angle and ellipticity obtained from models 2 and 3, in both cases for $\alpha_0 = \beta_0$ throughout the temperature range and for the case of $\beta_0 < \alpha_0$ (as given in figure 1S) at temperatures close to the phase transition (**Figure S5** and **S6**).

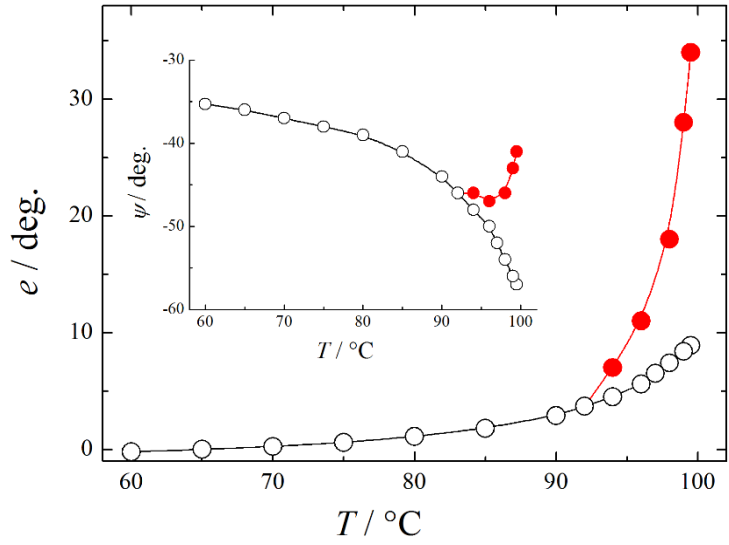

**Figure S5**. The ellipticity ($e$) and azimuthal angle ($\psi$), at the inset, as a function of temperature ($T$) calculated by using model 2. Black open circles: $\alpha_0 = \beta_0$; red solid circles: $\alpha_0 > \beta_0$.

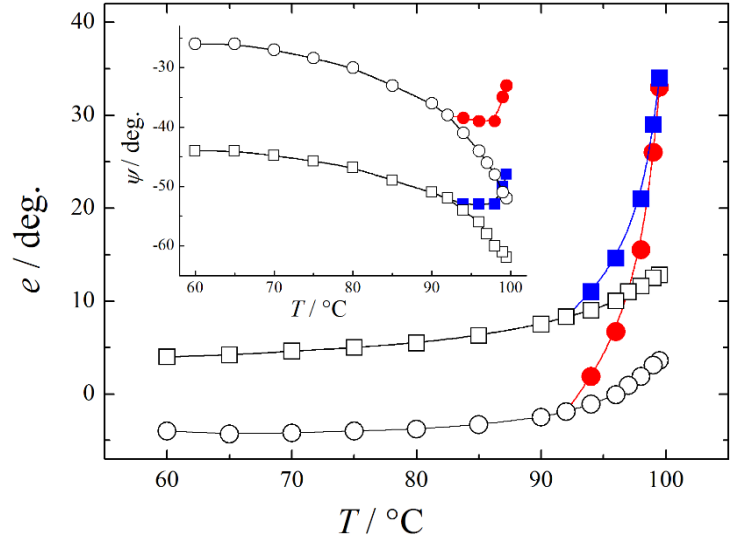

**Figure S6**. The ellipticity ($e$) and azimuthal angle ($\psi$), at the inset, as a function of temperature ($T$) calculated by using model 3. Black open circles: $q = 2q_0$ and $\alpha_0 = \beta_0$; black open squares: $q = -2q_0$ and $\alpha_0 = \beta_0$; red solid circles: $q = 2q_0$ and $\alpha_0 > \beta_0$; blue solid squares: $q = -2q_0$ and $\alpha_0 > \beta_0$.